\documentclass[aps,prl,twocolumn,superscriptaddress]{revtex4-1}

\usepackage{graphicx}
\usepackage{amsmath,amsfonts,amssymb}
\usepackage[colorlinks=true,linkcolor=blue,citecolor=blue]{hyperref}

\begin{document}

\title{Alternative types of molecule-decorated atomic chains in Au-CO-Au single-molecule junctions}

\author{Zolt{\'a}n~Balogh}
\affiliation{Department of Physics, Budapest University of Technology and Economics and MTA-BME Condensed Matter Research Group, Budafoki ut 8, 1111 Budapest, Hungary}
\author{P{\'e}ter Makk}
\affiliation{Department of Physics, Budapest University of Technology and Economics and MTA-BME Condensed Matter Research Group, Budafoki ut 8, 1111 Budapest, Hungary}
\author{Andr{\'a}s Halbritter}
\email{halbritt@mono.eik.bme.hu}
\affiliation{Department of Physics, Budapest University of Technology and Economics and MTA-BME Condensed Matter Research Group, Budafoki ut 8, 1111 Budapest, Hungary}

\begin{abstract}
We investigate the formation and evolution of Au-CO single-molecule break junctions. The conductance histogram exhibits two distinct molecular configurations, which are further investigated by a combined statistical analysis. According to conditional histogram and correlation analysis these molecular configurations show strong anticorrelations with each other and with pure Au monoatomic junctions and atomic chains. We identify molecular precursor configurations with somewhat higher conductance which are formed prior to single-molecule junctions. According to detailed length analysis two distinct types of molecule-affected chain formation processes are observed, and we compare these results to former theoretical calculations considering bridge and atop type molecular configurations where the latter has reduced conductance due to destructive Fano interference.
\end{abstract}

\maketitle

\section{Introduction}

The break junction method is widely used to establish single-molecule nanowires \cite{Elke_book,Venkata_review}. Along its controlled rupture a metallic wire thins down to atomic dimensions and finally breaks forming a nanometer-sized gap between the electrodes. This gap can be bridged by single molecules in a self-organized way. As the microscopic details of such molecular junctions can vary from experiment to experiment, a statistical analysis is necessary. The break junction method allows the statistical investigation of molecular junctions: by closing the junction the metallic electrodes can be reconnected, and afterwards by stretching and breaking the junction, new molecular junctions can be formed. 

Along the rupture conductance traces are recorded, i.e. the conductance of the breaking wire is measured as a function of electrode displacement. By repeating the break junction measurement several thousand times a statistical ensemble of conductance traces are collected, from which a conductance histogram can be plotted. Peaks in the histogram reflect the conductance of typical junction configurations, like single-atom contacts. After molecule dosing the formation of single-molecule nanowires is signalled by the appearance of new peaks in the histograms \cite{Elke_book,Venkata_review}.
However, to determine the details of molecular junction formation further analysis is required \cite{Elke_book, Venkata_review, doi:10.1021/nl060116e, Oren1, thermo1,Venkata_thermo, thermo2, thermo4, thermo5, thermo6, smit02, NDC, IETS, Oren3, IETS2, IETS3, yanson98, lowTH2_1, PtCO, Oren_chain, 2DCDH1, Venkata_2D, 2DCDH2, 2DCDH3, korrel1, korrel2, AgCO, Venkata1}. Important information can be obtained by further techniques like noise \cite{doi:10.1021/nl060116e, Oren1}, thermopower \cite{thermo1,Venkata_thermo} or inelastic spectroscopy \cite{smit02, NDC, IETS, Oren3} measurements, however, it has been recently realized that just by the advanced statistical analysis of the measured conductance traces essential information can be obtained. To this end several data analysis methods have been introduced, like plateaus' length histograms \cite{yanson98, lowTH2_1, PtCO, Oren_chain}, two-dimensional conductance-displacement histograms \cite{2DCDH1, Venkata_2D, 2DCDH2, PtCO} and correlation histograms \cite{korrel1, korrel2, AgCO, Venkata1}.

In this paper we investigate the formation of Au-CO-Au molecular junctions and show that in the presence of CO two new molecular configurations are formed. To identify them we carry out a statistical analysis going far beyond the simple conductance histogram approach. By combining conditional conductance histograms, two-dimensional correlation histograms (2DCH)\cite{korrel1, korrel2, AgCO} and two dimensional conductance-displacement histograms (2DCDH)\cite{2DCDH1, Venkata_2D, PtCO} we are able to narrow down the possible configurations formed during the breaking process. The comparison of our results with calculations of Refs.\cite{sim_real, sim_inf, sim_short} implies that the CO molecule can be either incorporated to gold atomic chains (bridge geometry), or can bind next to the chain (atop geometry). In the latter case the conductance of the atomic chain is substantially lowered as a result of destructive interference effect \cite{sim_real}.

\section{Results and Discussion}

\subsection{Conductance histograms and correlation analysis}

\begin{figure}
\begin{center}
\includegraphics[width=8.2cm,keepaspectratio]{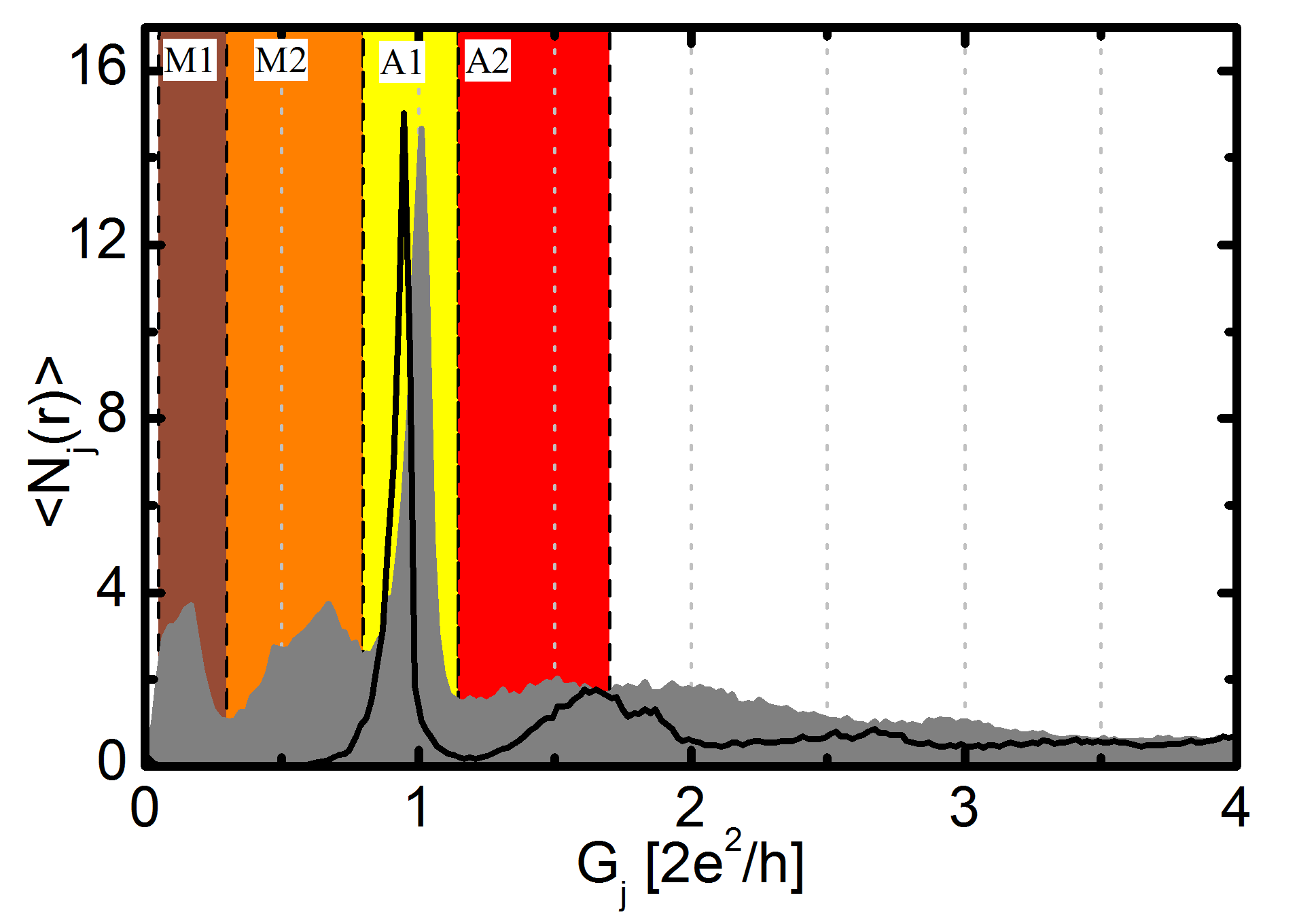}
\end{center}
\caption{\emph{Conductance histogram of Au junctions before (black) and after (grey) the dosing of CO molecules. The latter shows two new molecular peaks at $0.17$G$_{0}$ and $0.67$G$_{0}$. The color background shows the important conductance regions discussed in the text: $M1$ ($0.05-0.3$G$_{0}$, brown), $M2$ ($0.3-0.8$G$_{0}$, orange), $A1$ ($0.8-1.1$G$_{0}$, yellow) and $A2$ ($1.1-1.7$G$_{0}$, red). All histograms are normalized to the number of traces.}} \label{hist}
\end{figure}

The conductance histogram of clean gold junctions at T$=4.2\,$K is shown in Fig.~\ref{hist} with black curve. A sharp peak appears at $1$G$_{0}$, with smaller peaks at higher conductance, typical for low temperature measurements on Au junctions \cite{agrait03, PhysRevB.69.121411, yanson98}. It is known for gold that after stretching a monoatomic contact it does not necessarily break, but additional atoms can be pulled into the junction forming a nanowire with a single atom cross section and several atoms length \cite{yanson98,HRTEM2, smit01, calib, lowTH2_1}. The first peak in the histogram at $1$G$_{0}$ corresponds both to monoatomic junctions and to atomic chains.

After characterizing clean Au junctions CO molecules are dosed to the junction through a heated metallic tube (see details in Methods). The conductance histogram of Au-CO junctions is shown in Fig.~\ref{hist} with a grey area graph. The shape of the clean histogram is changed significantly after the dosing of CO: two new peaks appear in the region below $1$G$_{0}$ at $0.17$G$_{0}$ and $0.67$G$_{0}$. This histogram shows similarities to former measurements on Au-CO junctions \cite{CO1, Boer}, though in Ref.~\cite{CO1} only the molecular peak with higher conductance ($\sim 0.7$G$_{0}$), and in Ref.~\cite{Boer} the one with lower conductance $\sim 0.2-0.3$G$_{0}$ are reported, whereas in our measurement these two peaks coexist.

To simplify further analysis we define separate conductance regions marked by different colors in Fig.~\ref{hist}. The $M1$ ($0.05-0.3$G$_{0}$, brown) and $M2$ ($0.3-0.8$G$_{0}$, orange) regions correspond to the two molecular configurations; the region of monoatomic Au junctions and chains is marked by $A1$ ($0.8-1.1$G$_{0}$, yellow); whereas configurations with a bit higher conductance ($1.1-1.7$G$_{0}$, red) are labelled by $A2$. The latter region is attributed to \emph{precursor molecular configurations}, as discussed later and in Ref.~\cite{AgCO}. To gain further information about the nature of these configurations, and their relation to each other we use two dimensional correlation analysis, conditional histograms, and conditional two dimensional conductance-displacement histograms, which are constructed by the proper statistical analysis of the same conductance traces, that are used to plot the conductance histogram itself \cite{korrel1, korrel2, Venkata1, AgCO, PtCO}.

The two dimensional correlation histogram (2DCH) investigates the correlations between different conductance regions. Configurations that tend to appear together during the breaking process induce positive correlations in the 2DCH, whereas configurations excluding each other along a contact rupture induce negative correlations. It can also happen, that both configurations appear, but the lengths of the plateaus within the studied conductance bins exhibit distinct correlations. More details on the technique can be found in Refs.~\cite{korrel1, korrel2}.

\begin{figure}
\begin{center}
\includegraphics[width=8.2cm,keepaspectratio]{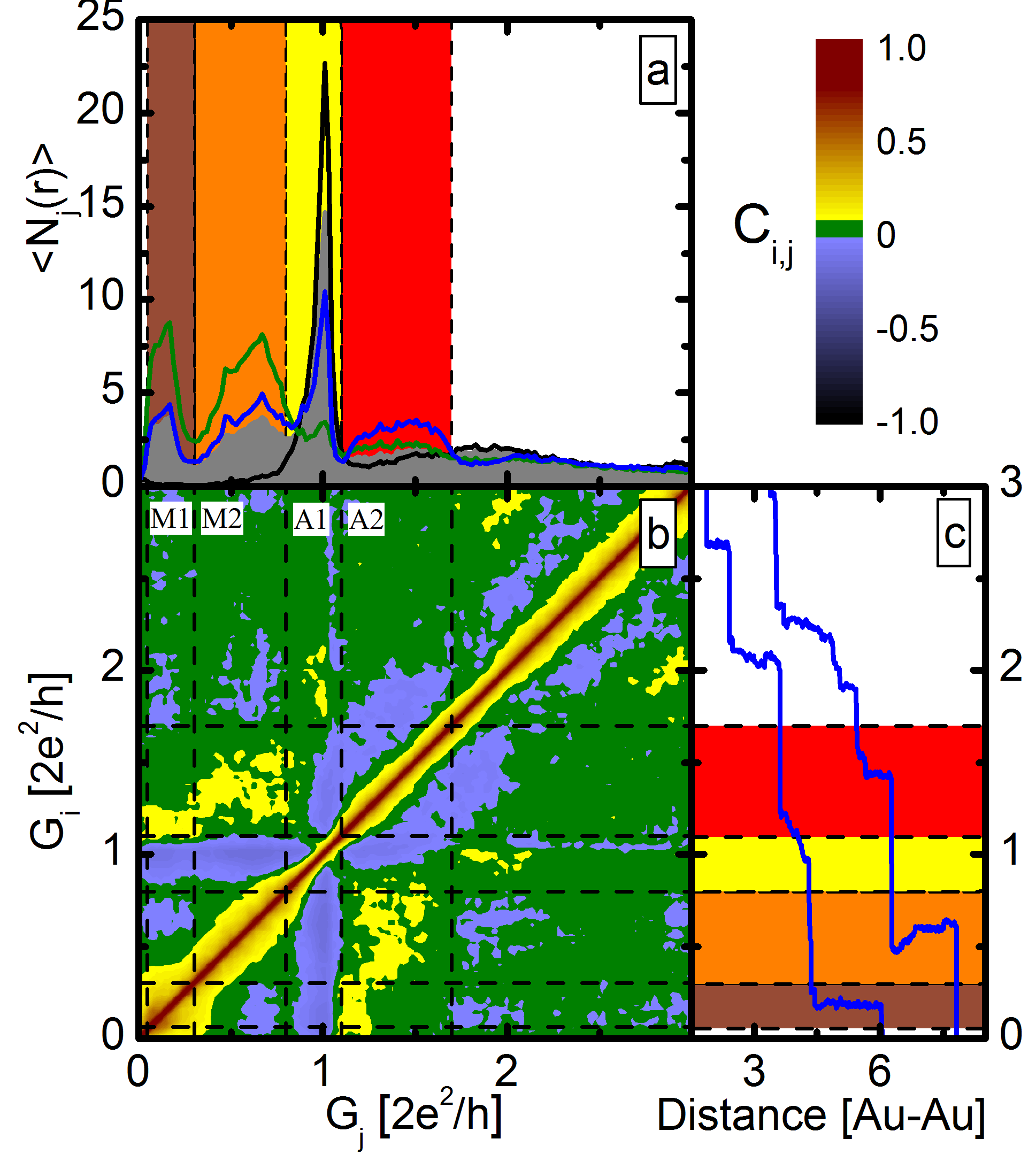}
\end{center}
\caption{\emph{(a) Conditional histograms for selected traces with larger than average length in different regions: $M1-M2$ (green), $A1$ (black) and $A2$ (blue). As a reference the grey area graph shows the histogram for all traces. (b) The 2D correlational histogram of Au-CO junctions. (c) Two demonstrative opening traces.}} \label{pull-correl}
\end{figure}

The 2DCH for Au-CO-Au junctions is shown in Fig.~\ref{pull-correl}b. Here the two axes correspond to the two conductances, and the value of the correlation function is shown by colors. Warm colors (yellow-red) correspond to positive correlations, cold colors (blue-black) to negative and green marks configurations which are independent within the resolution of the method.

The 2DCH in Fig.~\ref{pull-correl}b shows clear anti-correlation between the both molecular configurations ($M1-M2$) and the single atomic configuration ($A1$). However, the region above $1$G$_{0}$ ($A2$), shows positive correlation with both molecular configurations $M1-M2$. Between the two molecular configurations $M1$ and $M2$ an anticorrelation is observed.

Whereas 2DCHs are useful to make a two dimensional map of the relevant correlation effects, to gain a more quantitative picture about the nature and strength of correlations between different regions it is also useful to investigate conditional histograms.
Conditional conductance histograms are conductance histograms for breaking traces selected according to a predefined condition. Here we focus on conductance traces, which are long enough in a predefined conductance region. To construct these histograms, we have determined the average length of breaking traces in the predefined region and selected the conductance traces longer than average. Conditional histograms are normalized to the number of traces included, therefore direct comparison of them is possible. We note here, that the selected data sets obtained by this technique are not distinct, because the length of a given trace can be longer than the average in more regions (positive correlation). More details on the method can be found in Refs.\cite{korrel2, PtCO, AgCO}.

Conditional conductance histograms selected for different conductance regions are shown in Fig.~\ref{pull-correl}a. As a reference, the grey area curve shows the histogram for all traces (5000 curves). The conditional conductance histogram constructed for the $A1$ region (black) shows the full suppression of the peaks in the molecular region $M1-M2$. These selected traces correspond to the case, when pronounced (long) $\sim1$G$_0$ plateaus are observed. According to the conditional histogram these traces do not show molecular configurations, i.e. the corresponding negative correlation in the 2DCH between $A1$ and $M1-M2$, is a strong feature: the $A1$ and $M1-M2$ configurations practically exclude each other. Similar feature is seen from the other side as well: the conditional histogram selected for the molecular region $M1-M2$ (green line) exhibits the almost complete suppression of the peak at $1$G$_0$. This conditional histogram shows an additional interesting feature: the enhanced weight in the $A2$ region compared to the total histogram, which was also reflected by the positive correlation between $M1-M2$ and $A2$ in the 2DCH (And similarly the blue conditional histogram for the $A2$ region exhibits enhanced molecular peaks). All these imply, that Au-CO-Au single-molecule junctions are usually not preceded by clean Au monoatomic contacts or atomic chains, rather a so-called precursor configuration with higher conductance (A2 region), similarly to our previous report on Ag-CO junctions \cite{AgCO}. We interpret these precursor configurations by the binding of a CO molecule to the side of a single-atom contact, which opens additional conductance channel(s) compared to the single-channel transport in pure Au monoatomic contacts and atomic chains. Fig.~\ref{pull-correl}c shows two example traces with a molecular plateau preceded by a precursor configuration.

\begin{figure}
\begin{center}
\includegraphics[width=8.2cm,keepaspectratio]{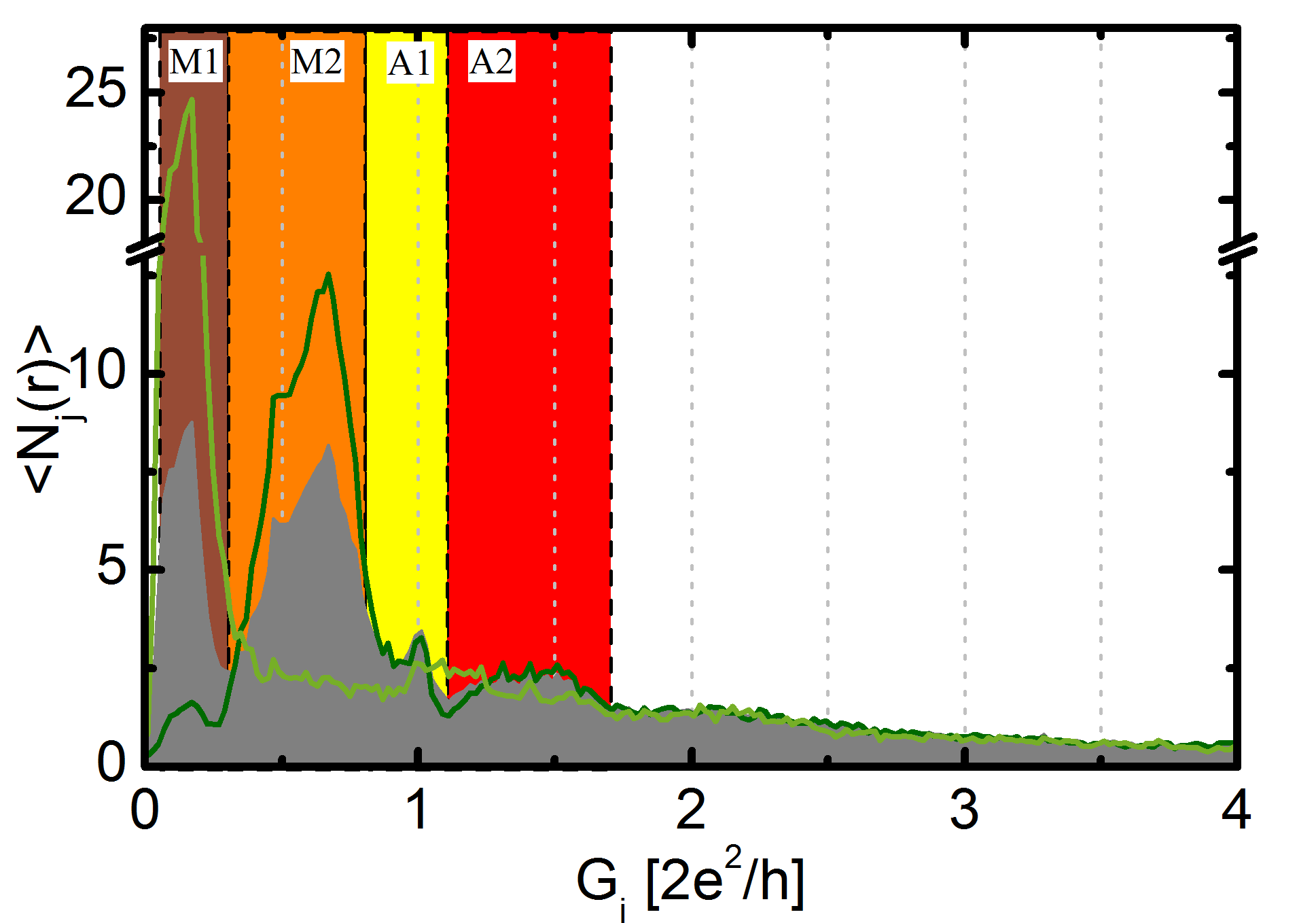}
\end{center}
\caption{\emph{Conditional histograms based on double selection. First the traces are selected for which the molecular plateau in the entire $M1-M2$ region is longer than average, i.e. the traces corresponding to pronounced molecular junctions. The conditional histogram for these traces (grey area graph) is the same as the green line in Fig.~\ref{pull-correl}a. As a next step we select traces from this reduced dataset, for which the M1 (M2) region is longer than average, respectively. The light and dark green curves show the corresponding conditional histograms for the $M1$ and $M2$ region, respectively.}} \label{hist_molsel}
\end{figure}

To investigate the relation of the two molecular configurations to each other we have selected the opening traces, where the molecular configurations have been formed (green curve in Fig.~\ref{pull-correl}a). This accounts to around $\sim 40\%$ of all the breaking traces. Afterwards we have constructed the conditional conductance histogram for this reduced dataset for region $M1$ and $M2$ separately, which is shown in Fig.~\ref{hist_molsel}. In the conditional histogram selected for the $M1$ region (green) no peak appears in the $M2$ region, and similarly the conditional histogram for the $M2$ configuration (dark green) shows an almost complete suppression of the peak in the $M1$ region. These observations point to a strong anti-correlation between the two molecular configurations: if one molecular configuration is pronounced, the other is practically absent. This is in agreement with the anticorrelation of regions $M1$ and $M2$ shown by the blue spots in the 2DCH.

\subsection{Conditional two-dimensional conductance-displacement histograms}

So far we have investigated the appearance of different configurations and their relation, but the shape of the individual traces has been discarded in this analysis. Two typical opening traces are shown in Fig.~\ref{pull-correl}c, which are quite different in nature. Instead of the investigation of the individual traces we investigate 2D conductance-displacement histograms(2DCDH) \cite{2DCDH1, Venkata_2D, 2DCDH2, PtCO}. This method provides information about the conductance and the corresponding breaking length at the same time.

To construct the 2DCDH a reference conductance, $G^\mathrm{ref}$ is chosen and the traces are shifted along the displacement axis, such, that all the curves have this reference conductance at the zero point of the displacement axis. Afterwards, by dividing the displacement and conductance axis into discrete bins, a 2D histogram is constructed, where the number of counts is shown by colors.

We have chosen $1.7\,$G$_{0}$ for reference value, which is the upper boundary of the precursor region, $A2$. The electrode separation was calibrated by forming atomic chains in clean gold nanojunctions and by measuring the distance of the peaks in the plateaus' length histogram \cite{calib, lowTH2_1, PtCO}. Further on we scale the displacement in the units of the Au-Au distance in pure monoatomic chains, which is around $2.6$\AA \cite{calib}. We have verified this calibration by measuring tunneling current as a function of electrode displacement.

\begin{figure}
\begin{center}
\includegraphics[width=8.2cm,keepaspectratio]{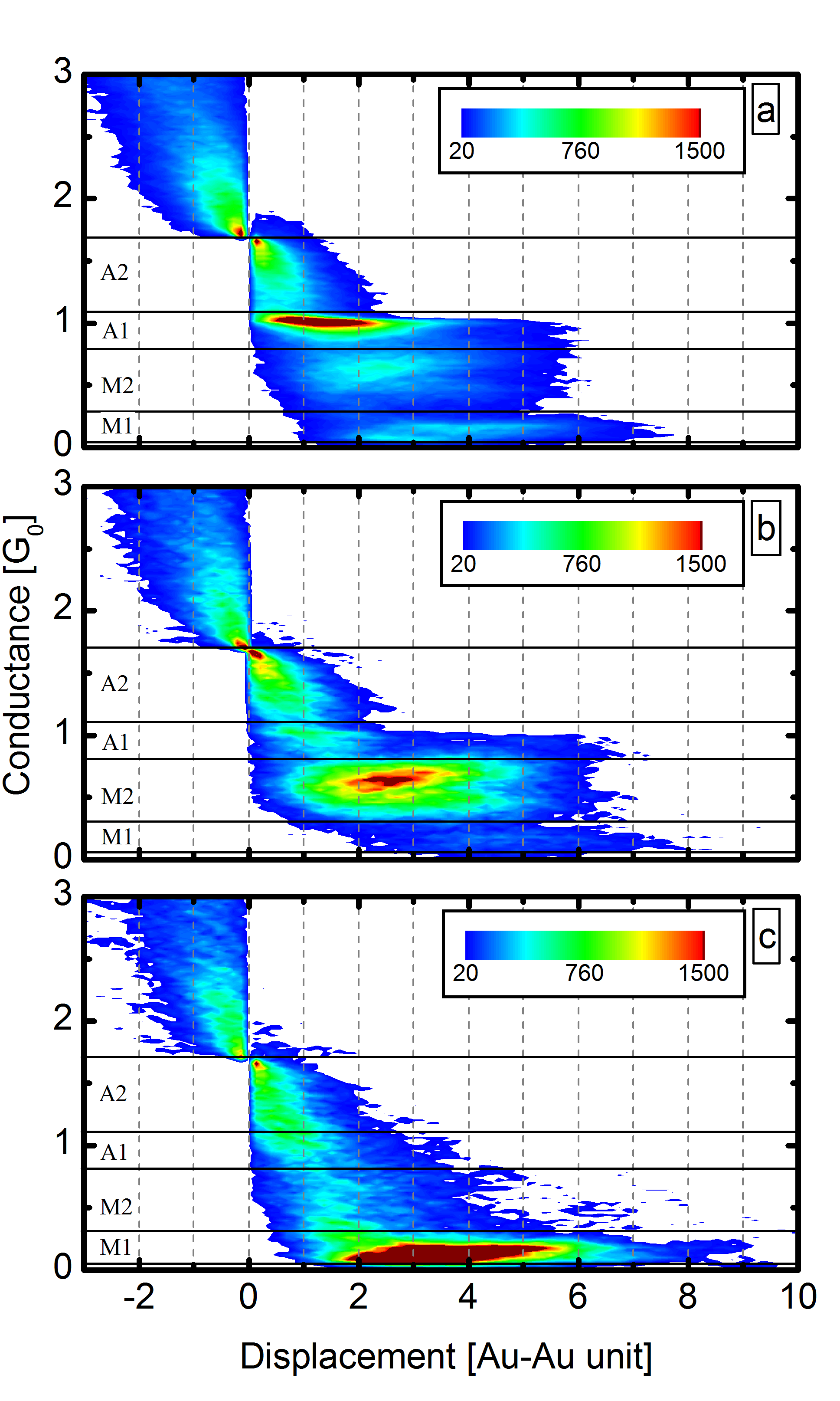}
\end{center}
\caption{\emph{(a) Two-dimensional conductance-displacement histogram for all the measured traces, showing the appearance of the same configurations as the grey curve in Fig.~\ref{hist}.  (b-c) The 2D conductance-displacement histograms for the same traces that were used to plot the light/dark green, double selected conditional histograms in Fig.~\ref{hist_molsel}. Panels (b)/(c) correspond to molecular traces selected for the $M2$/$M1$ region, respectively. The electrode-displacement is given in Au-Au distance unit, i.e. the distance of neighbor Au atoms in clean atomic chains. The 2DCDHs are normalized to the number of included traces.}} \label{Trace_peak}
\end{figure}

In Fig.~\ref{Trace_peak}a the 2DCDH is shown for all traces, demonstrating that the $A1$, $M1$ and $M2$ confugurations all have extended plateaus  indicating atomic chain formation in all regions. As all the traces are included in this plot, different types of breaking traces can not be distinguished in the 2DCDH. To resolve the typical trajectories, we have investigated the 2DCDH for separate sets of curves.

Figure~\ref{Trace_peak}b-c shows the 2DCDHs for the same double-selected datasets that were used to plot the light and dark green conditional histograms in Fig.~\ref{hist_molsel}, i.e for molecule-affected opening traces longer than average in the $M1$ or $M2$ region. Both plots exhibit an extended plateau in the region of the selection with up to $5-6\,$Au-Au unit length, and a substantial suppression of the other two regions. This points to two distinct types of molecule assisted chain formation processes: either a chain with $M1$ or $M2$ conductance is formed, but there is no significant transition between these two. We emphasize that this is in clear contrast to our previous observations on Pt-CO junctions \cite{PtCO}, where Pt atomic chains can be pulled either by a perpendicular or a parallel CO molecule sitting in the chain, and the molecule frequently rotates from the perpendicular to the parallel orientation along the chain formation process, i.e. a clear transition is observed between the corresponding two conductance intervals.

\subsection{Final configuration analysis}

\begin{figure}
\begin{center}
\includegraphics[width=8.2cm,keepaspectratio]{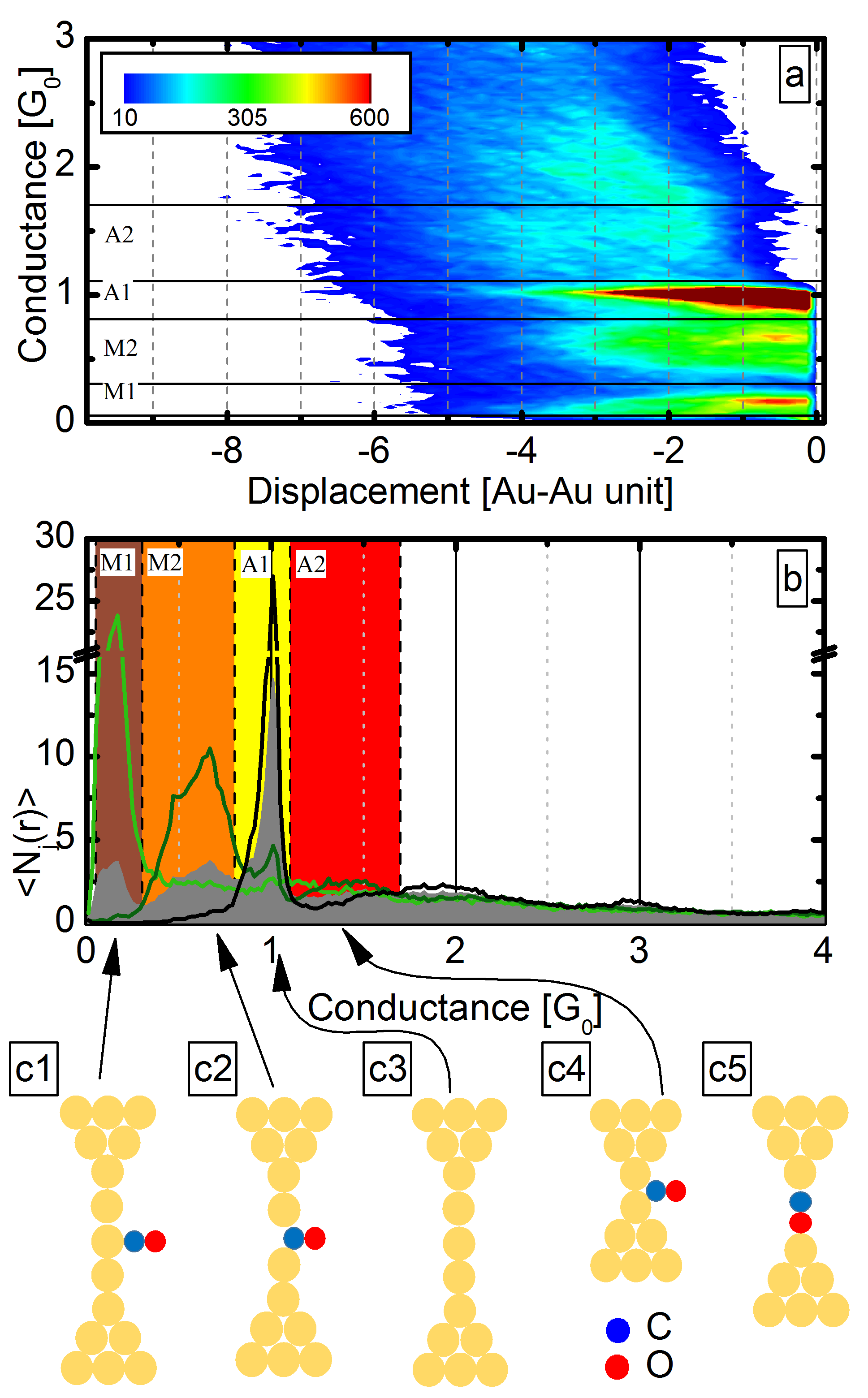}
\end{center}
\caption{\emph{(a) The 2DCDH of all traces with $G^\mathrm{ref}=0.01\,$G$_{0}$. This histogram shows that the junction can break from all the three regions ($M1-M2$, $A1$). (b) The histograms of selected traces according to the final configuration selection for the $M1$ region (light green), $M2$ region (dark green) and the $A1$ region (black).  (c1-c5) Configurations that can be formed during the junction opening: (c1)/(c2) Au atomic chain with a CO molecule bound in an atop/bridge geometry, respectively. (c3) Pure Au atomic chain. (c4) Precursor molecular configuration. (c5) A parallel CO molecule is incorporated in the chain. (According to our arguments the latter is not observed in the present experiments.)}} \label{Trace_break}
\end{figure}

We note here, that particular features are enhanced in the above 2DCDH analysis: for example since we have selected long traces for the defining regions all the displayed curves in this region are longer than average.  Also the choice of reference conductance affects the shape of the 2DCDH map. Therefore the 2DCDH of all traces with different $G^\mathrm{ref}$ is shown in Fig.~\ref{Trace_break}a. Here all the curves are fitted together at the breaking point, which we define as $G^\mathrm{ref}=0.01$G$_{0}$. We can observe that after extended plateaus in the $A1$, $M2$ or $M1$ region the junction can break from all these configurations.

Similarly to our study on Pt-CO junctions\cite{PtCO} it is useful to select the traces according to their final configuration. A trace is considered to break from one of the three contact configurations ($A1$, $M2$ or $M1$) if $70$\%
of the final $30$ data points ($\approx 1$\AA) before the rupture is within the corresponding conductance region. According to this selection respectively $42$\%, $27$\% and $14$\% of all traces break from the $A1$, $M2$ or $M1$ region, and for the remaining $17$\% of the traces our algorithm does not identify a clear final configuration.

The conditional histogram for the selected traces according to their final configuration is shown in Fig.~\ref{Trace_break}b. As a reference the grey area graph shows the histogram of all the recorded traces. The final configuration histogram for the $M2$ region (dark green curve) shows a strong suppression of the $M1$ and $A1$ region, and an enhancement of the $A2$ region; whereas the final configuration histogram of the $M1$ region (light green) shows a strong suppression of the $M2$ and $A1$ region, and it also shows an enhancement of the $A2$ region, which support our previous finding: on one trace only one of the two molecular configurations is pronounced. This analysis also shows, that the $M2$ molecular configuration has a precursor configuration centered at the middle of the $A2$ region, whereas the precursor configuration of the $M1$ configuration has somewhat lower conductance, centered rather at the border of the $A1$ and $A2$ regions. The final configuration histogram for the $M2$ region also shows a minor peak around $1\,$G$_0$, demonstrating that the $M2$ configuration is not always preceded by the $A2$ precursor molecular configuration, but occasionally it is preceded by a pure Au monoatomic contact. The latter observation is also supported by the 2DCDH in Fig.~\ref{Trace_peak}b.

\subsection{Discussion}

Our analysis showed that the junctions can break from three different configurations ($A1$, $M1$, $M2$) and on one trace only one out of these three configurations is pronounced with a length up to $5-6\,$Au-Au unit. This means that three distinct chain formation processes are observed: (i) pure Au monoatomic chains; (ii) molecule affected chains with $M2$ conductance; (iii) molecule affected chains with $M1$ conductance. According to our analysis there is no significant transition between these three types of chains.

To identify the $M1$ and $M2$ molecular configurations we compare our results to the ab initio simulations in Refs.~\cite{sim_real,sim_inf}. The conductance of the second molecular peak ($M2$) coincides with the calculated conductance of a bridge-like molecular configuration ($0.6-0.7$G$_{0}$), where the CO molecule sits between two neighbor Au chain atoms in a perpendicular direction with respect to the contact axis (see Fig.~\ref{Trace_break}c2), similarly to our observations on Pt-CO junctions \cite{PtCO}.

The $M1$ molecular configuration could be related to another configuration, where the CO molecule sits between two Au atoms in a parallel direction (see Fig.~\ref{Trace_break}c5). However, it is reasonable to assume that before this configuration a perpendicular configuration is formed, which is then rotated to the parallel orientation, as it was observed in Pt-CO molecule-decorated chains \cite{PtCO}. In the present experiments no transition between the two molecular configurations is observed, therefore we do not favor the interpretation of the $M1$ configuration as a parallel CO molecular junction.

As an alternative we consider the so-called atop geometry reported in Refs.~\cite{sim_real,sim_inf}, where the CO molecule is not wedged in between two Au atoms, but it binds to the side of a single Au atom of a gold atomic chain (see Fig.~\ref{Trace_break}c1). This geometry is interesting, since one would expect perfect transmission for the undistorted Au atomic chain, but the CO molecule binding to the side induces a strong reduction of the conductance due to Fano-like destructive interference effect. The calculated conductance of this geometry ($0.08$G$_{0}$)\cite{sim_real,sim_inf} agrees with the conductance of the $M1$ region. We favor this interpretation, as it can also account for the absence of transition between the molecule affected chains with $M1$ and $M2$ conductance.

Similarly to our previous study on Ag-CO junctions the $A2$ precursor configuration is considered as a CO molecule bound to the side of a dimer Au junction (see Fig.~\ref{Trace_break}c4).

We also note that the $M2$ configuration exhibits a clear positive slope of the conductance plateau (see Fig.~\ref{Trace_peak}b and also the second sample trace in Fig.~\ref{pull-correl}c), which agrees with Refs.~\cite{sim_real,CO1}. 

\section{Conclusions}

In this paper we have investigated the formation and evolution of Au-CO single-molecule junctions. To get more insight into the junction formation we used different statistical analysis methods, like correlational analysis, 2D conductance-displacement histograms, and conditional histograms. The combination of these techniques can reveal information far beyond simple conductance histogram measurements.
We have observed the formation of two molecular configurations, which could be stretched through several atom-atom distances, forming molecule-decorated atomic chains. The evolution of Au-CO atomic chains is in clear contrast to our previous report on Pt-CO molecule affected chain formation. In the latter case the molecular configuration with higher conductance (perpendicular CO) is transformed to the molecular configuration with lower conductance (parallel CO) along the chain formation. In the present experiments, however, no transition is observed between the two distinct types of molecule-decorated chains. This behaviour is naturally explained if the low conductance molecular configuration is not a derivative of the high conductance one, rather it is a completely different structural arrangement. The two simulated molecular configurations in Refs.~\cite{sim_real, sim_inf} might account for our observations, as the higher conductance of the bridge geometry and Fano interference suppressed lower conductance of the structurally different atop geometry coincide with the experimentally observed peak positions. A precursor molecular configuration was also observed, from which the molecular junctions are likely to be formed.

\section{Methods}
The measurements were performed with a self-designed MCBJ setup at liquid helium temperature and the CO molecules were dosed with a home-made vacuum-system from a high purity container through a heated tube. The detailed description of our experimental technique is introduced in our previous publication \cite{AgCO}.

We have performed our measurements on one break junctions sample that was measured for several weeks as follows. In order to exclude unwanted contamination we have a very strict protocol for this type of measurement. First we bake out and pump the sample holder for approximately one week. Afterwards we cool down the sample holder and record clean histograms. As a next step we precisely mimic all steps of molecule dosing (heating of the dosing tube, opening the shutter and the proper valves, etc.) except that the gas container remains closed and we check that the clean metal histogram is preserved. After the real dosing of the molecules, new peaks appeared and were present at moderate bias voltage ($\approx 30-50\,$mV) for a long time ($10'000-20'000$ opening-closing cycle) without further dosing. Then we could permanently desorb the molecules by high bias voltage ($\geq 100\,$mV) histogram measurement. With this step clean Au histogram was reestablished, and no molecular peaks were observed even if the bias voltage was reduced back ($\approx 30-50\,$mV), so a repeated dosing was necessary to see molecular peaks again. In the present study three independent data sets were acquired by permanent desorption of the CO molecules and repeated dosing, which show reproducible data.

The displacement of the electrodes was calibrated prior to molecule dosing using both plateaus' length histograms and conductance traces in the tunneling region. To ensure proper cleanliness during the molecular measurements the turbomolecular pump was continuously running. This has introduced $<1\,$\AA-scale vibrations to the junction, which resulted in the appearance of false peaks in the plateaus' length histograms of molecular junctions, thus we have avoided drawing conclusions from molecular plateaus' length histograms.

\begin{acknowledgements}
We acknowledge the financial support from the Hungarian Scientific Research Fund, OTKA K105735 research grant.
\end{acknowledgements}
\bibliographystyle{prsty}
\bibliography{AuCO}

\vspace{3cm}
This article is published in full length in \textit{Beilstein Journal of Nanotechnology}
\textbf{2015}, \textit{6}, 1369–1376.

\end{document}